\documentstyle[12pt]{article}
\evensidemargin\oddsidemargin
 \catcode`@=11
\def\@cite#1#2{\unskip\nobreak\relax
    \def\@tempa{$\m@th^{\hbox{\the\scriptfont0 #1}}$}%
    \futurelet\@tempc\@citexx}
\def\@citexx{\ifx.\@tempc\let\@tempd=\@citepunct\else
    \ifx,\@tempc\let\@tempd=\@citepunct\else
    \let\@tempd=\@tempa\fi\fi\@tempd}
\def\@citepunct{\@tempc\edef\@sf{\spacefactor=\the\spacefactor\relax}\@tempa
    \@sf\@gobble}

\def\citenum#1{{\def\@cite##1##2{##1}\cite{#1}}}
\def\citea#1{\@cite{#1}{}}

\newcount\@tempcntc
\def\@citex[#1]#2{\if@filesw\immediate\write\@auxout{\string\citation{#2}}\fi
  \@tempcnta\z@\@tempcntb\m@ne\def\@citea{}\@cite{\@for\@citeb:=#2\do
    {\@ifundefined
       {b@\@citeb}{\@citeo\@tempcntb\m@ne\@citea\def\@citea{,}{\bf ?}\@warning
       {Citation `\@citeb' on page \thepage \space undefined}}%
    {\setbox\z@\hbox{\global\@tempcntc0\csname b@\@citeb\endcsname\relax}%
     \ifnum\@tempcntc=\z@ \@citeo\@tempcntb\m@ne
       \@citea\def\@citea{,}\hbox{\csname b@\@citeb\endcsname}%
     \else
      \advance\@tempcntb\@ne
      \ifnum\@tempcntb=\@tempcntc
      \else\advance\@tempcntb\m@ne\@citeo
      \@tempcnta\@tempcntc\@tempcntb\@tempcntc\fi\fi}}\@citeo}{#1}}
\def\@citeo{\ifnum\@tempcnta>\@tempcntb\else\@citea\def\@citea{,}%
  \ifnum\@tempcnta=\@tempcntb\the\@tempcnta\else
   {\advance\@tempcnta\@ne\ifnum\@tempcnta=\@tempcntb \else \def\@citea{--}\fi
    \advance\@tempcnta\m@ne\the\@tempcnta\@citea\the\@tempcntb}\fi\fi}
\catcode`@=12

\renewenvironment{thebibliography}[1]
 {\begin{list}{\arabic{enumi}.}
    {\usecounter{enumi} \setlength{\parsep}{0pt}
     \setlength{\itemsep}{3pt} \settowidth{\labelwidth}{#1.}
     \sloppy
    }}{\end{list}}

\newcommand{\alt}{\mathrel{\raisebox{-.6ex}{$\stackrel{\textstyle<}{\sim}$}}}
\newcommand{\agt}{\mathrel{\raisebox{-.6ex}{$\stackrel{\textstyle>}{\sim}$}}}
\let\ltap=\alt \let\gtap=\agt

\def\overlay#1#2{\ifmmode \setbox 0=\hbox {$#1$}\setbox 1=\hbox to\wd 0{\hss
$#2$\hss }\else \setbox 0=\hbox {#1}\setbox 1=\hbox to\wd 0{\hss #2\hss }\fi
#1\hskip -\wd 0\box 1}

\def\nv#1 {\noalign{\vskip#1pt}}

\def\abstract#1{\begin{center}\Large\bf Abstract\end{center}
{\narrower\small #1\par}}

\def\gev{{\rm\,GeV}}

\def\L{{\cal L}}

\def\pb{{\rm\,pb}}

\def\b#1,{{\bf #1}}
\def\zero{\phantom0}

\renewcommand{\labelenumi}{(\alph{enumi})}

\begin{document}

\font\fortssbx=cmssbx10 scaled \magstep2
\hbox to \hsize{
\hskip.5in \raise.1in\hbox{\fortssbx University of Wisconsin - Madison}
\hfill\vbox{\hbox{\bf MAD/PH/765}
            \hbox{June 1993}} }

\vspace{.35in}

\begin{center}
{\Large\bf SUPERSYMMETRY PHENOMENOLOGY\footnote{Talk presented by V.~Barger at
the {\it HARC SUSY Workshop}, Texas, April 1993}}\\[.3in]
{\large V.~Barger$^{\,a}$  and R.J.N.~Phillips${\,^b}$}\\[.2in]
\it
$^a$Physics Department, University of Wisconsin, Madison, WI 53706, USA\\
$^b$Rutherford Appleton Laboratory, Chilton, Didcot, Oxon OX11 0QX, UK
\end{center}

\renewcommand{\LARGE}{\Large}
\renewcommand{\Huge}{\Large}

\vspace{.25in}

\abstract{
We review some recent results and future prospects in the phenomenology
of Supersymmmetry.  We discuss the searches for superpartner states,
the searches for Higgs bosons in the minimal SUSY model, and additional
parameter constraints arising in SUSY-GUT models.}

\section{Introduction}

   Supersymmetry (SUSY) is a vital element in present thinking about
fundamental particle physics.  At this Conference it certainly needs no
formal introduction nor motivation; we simply emphasise here two
important aspects --- it has perturbatively calculable consequences and
can be experimentally tested.  This talk will be about the associated
phenomenology.

   Experimental evidence for SUSY could come in various forms, such as
\begin{enumerate}

\item discovery of one or more superpartners,
\item discovery of a neutral Higgs boson with non-standard properties
    and/or a charged Higgs boson,
\item discovery of $p \to K \nu$ decay,
\item discovery that dark matter consists of heavy neutral particles.
\end{enumerate}
This talk will address areas (a) and (b) above.   We shall discuss the
present status and future prospects of searches for superpartner
particles, searches for Higgs bosons in the minimal SUSY model (MSSM)
and additional parameter constraints that come from SUSY-GUT studies.
We leave other phenomenological topics like proton decay, dark matter,
$R$-parity breaking, etc., to other speakers.

\section{Experimental SUSY Signatures}

The MSSM conserves a multiplicative $R$-parity: $R_p=(-1)^{2S+L+3B}$,
where $S,\,L,\,B$ are spin, lepton number, baryon number.  Standard
Model (SM) particles have $R=1$; their superpartners have $R=-1$. Hence heavy
sparticles must decay to lighter sparticles, while the lightest
sparticle (LSP) is stable.  If the LSP has no strong or electromagnetic
interactions, as seems likely since it has defied detection for so long,
it will carry away ``missing" energy and momentum --- a vital SUSY
signature.  The LSP will also be a candidate for cosmological dark matter;
this focuses extra interest on the $R_p$-conserving case.

Grand Unified Theories (GUTs) are essential for SUSY phenomenology to economize
parameters. A minimal set of GUT parameters with soft SUSY breaking consists of
the gauge and Yukawa couplings $g_i$ and $\lambda_i$, the Higgs mixing mass
$\mu$, the common gaugino mass at the GUT scale $m_{1/2}$, the common scalar
mass at the GUT scale $m_0$, and two parameters $A,B$ that give trilinear and
bilinear scalar couplings. At the weak scale, the gauge couplings are
experimentally determined. The Higgs potential depends upon $m_0,\mu,B$ (at
tree level) and $m_{1/2},A,\lambda_t,\lambda_b$ (at one loop). After minimizing
the Higgs potential and putting in the measured $Z$ and fermion masses, there
remain 5 independent parameters, e.g.\ $m_t,\tan\beta,m_0,m_{1/2},A$.

Beside the SM gauge bosons and fermions, the MSSM spectrum contains Higgs
bosons $(h,H,A,H^\pm)$, gluinos $(\tilde g)$, squarks $(\tilde q)$, sleptons
$(\tilde\ell^\pm)$, charginos ($\tilde W_i^\pm, i=1,2$; mixtures of winos and
charged higgsinos), neutralinos ($\tilde Z_j, j=1,2,3,4$; mixtures of zinos,
photinos and neutral higgsinos). An alternate notation is $\tilde\chi_i^\pm$
for $\tilde W_i^\pm$ and $\tilde\chi_j^0$ for $\tilde Z_j$. Figure~1
illustrates the evolution of the SUSY mass spectrum from the GUT
scale~\cite{ross,rosrob}. The running masses are plotted versus
$\mu$ and the physical value occurs approximately where the running mass
$m=m(\mu)$ intersects the curve $m=\mu$. For the Higgs scalar $H_2$, the
mass-square becomes negative at low $\mu$ due to coupling to top; in this
region we have actually plotted $-|m(\mu)|$. Negative mass-square parameter is
essential for spontaneous symmetry-breaking; here it is achieved by radiative
effects. The running masses for the gauginos $\tilde g,\tilde W,\tilde B$ are
given by
\begin{equation}
M_i(\mu)=m_{1/2}\,{\alpha_i(\mu)\over\alpha_i(M_G)} \,,
\end{equation}
where $i$ labels the gauge symmetry; this applies before mixing with higgsinos
to obtain the chargino and neutralino mass eigenstates. In  Fig.~1 the squarks
are heavier than the gluinos, but the opposite ordering $m_{\tilde q}<
m_{\tilde g}$ is possible in other scenarios.
Sleptons, neutralinos and charginos are lighter than squarks and gluinos in
general.
The usual soft SUSY-breaking mechanisms preserve the
gauge coupling relations (unification) at $M_G$.

   In order that SUSY cancellations shall take effect at low mass
scales as required, the SUSY mass parameters are expected to be
bounded by
\begin{equation}
      m_{\tilde g},\, m_{\tilde q},\, |\mu|,\, m_A \alt 1\mbox{--2 TeV}\, .
\label{m bound}
\end{equation}

\medskip
\begin{center}
{\small Fig.~1. Typical RGE results for spartner masses~\cite{ross}.}
\end{center}
\medskip

\noindent
The other parameter $\tan\beta$ is effectively bounded by
\begin{equation}
          1  \alt    \tan\beta  \alt    65    \;, \label{tanbetabound}
\end{equation}
where the lower bound arises from consistency in GUT models and the
   upper bound is a perturbative limit (following Ref.~3). 
Proton decay gives  $\tan\beta<85$~\cite{lopez2,hisano}.

      At LEP\,I, sufficiently light SUSY particles would be produced
through their gauge couplings to the $Z$.  Direct searches for
SUSY particles at LEP give
\begin{equation}
      m_{\tilde q},\, m_{\tilde \ell},\,m_{\tilde\nu},\,
 m_{\tilde W_1}\agt\mbox{40--45 GeV}\;.
\end{equation}
There are also indirect limits on $Z$ decays to SUSY particles\cite{bt},
\begin{eqnarray}
\Delta\Gamma_Z({\rm SUSY}) &<& 25\rm~MeV \,,\\
\Gamma(Z\to\tilde Z_1 \tilde Z_1:\ \rm invisible) &<& \rm 17~MeV\,,\\
B(Z\to\tilde Z_1 \tilde Z_1) &<& \rm few \times 10^{-5} \,.
\end{eqnarray}
It is convenient to express the excluded regions in terms of
the parameters that determine the gaugino masses, i.e.\ in the
$(\mu, m_{1/2})$ or $(\mu, m_{\tilde g})$ plane: see Fig.~2.

 Hadron colliders can explore much higher energy ranges, producing squarks and
gluinos strongly. At the Tevatron, with luminosity 25~pb$^{-1}$
expected in 1993,  about 100 events would be expected
for each of the channels $\tilde g\tilde q$ and $\tilde q\tilde q$ for
$m_{\tilde q} \sim m_{\tilde g} \sim 200$~GeV, reaching well beyond the
LEP range\cite{bt,bbdkt,btw}.

SUSY particle signatures depend on their decays.
The most distinctive  is the missing
energy and momentum of the undetected LSP  (usually
assumed to be the lightest neutralino $\tilde Z_1$)  in
all SUSY decay chains with $R$-parity conservation.  At hadron colliders
one can only do book-keeping on the missing transverse
momentum denoted $\overlay/p_T$, where the missing momenta of both LSPs are
added vectorially.  The LSP momenta and hence the magnitude
of $\overlay/p_T$ depend on the decay chains.

\medskip
\begin{center}
\parbox{5.5in}{\small Fig.~2: MSSM parameter regions accessible to LEP,
from Ref.~7. 
Solid curves surround
regions excluded by LEP\,I data, dot-dashed curves show the potential
reach of LEP\,II, and the dashed curve is the Tevatron CDF limit for
$\tan \beta = 2$.}
\end{center}
\medskip

 If squarks and gluinos are rather light ($m_{\tilde g},m_{\tilde q}\alt
50$~GeV), their dominant decay mechanisms are strong decays to each other or
decays to the LSP:
\begin{eqnarray}
\left.\begin{array}{l}
      \tilde q \to q \tilde g\\
\tilde g \to q \bar q \tilde Z_1
      \end{array} \right\}
&& {\rm if}\ m_{\tilde g} < m_{\tilde q} \,, \label{light a}\\
\left.\begin{array}{l}
           \tilde g\to q \tilde q \phantom{Z_1} \\
\tilde q\to  q \tilde Z_1  \phantom{q}
\end{array}\right\}
&&    {\rm if}\ m_{\tilde q} < m_{\tilde g} \,. \label{light b}
\end{eqnarray}
   In these cases the LSPs carry a substantial fraction of the available
   energy and $\overlay/p_T$ is correspondingly large.  Assuming such decays
and small LSP mass, the present 90\%~CL experimental bounds from UA1 and UA2
   at the CERN $p$-$\bar p$ collider ($\sqrt s = 640$~GeV) and from CDF at the
   Tevatron ($\sqrt s = 1.8$~TeV) are~\cite{uacdf}
\[
\vbox{\tabskip2em\halign{#\hfil&&$#$\hfil\cr
                    &  \hfil  m_{\tilde g}  &  \hfil  m_{\tilde q}\cr
        UA1 (1987)      &     >  \zero53\rm\ GeV  &  >  \zero45\rm\ GeV \cr
        UA2 (1990)      &     >  \zero79          &  >  \zero74 \cr
        CDF (1992)      &     >      141          &  >      126 \cr}}
\]
   The limits rise if the squark and gluino masses
   are assumed to be comparable.

   For heavier gluinos and squarks, many new decay channels are open,
such as decays into the heavier gauginos:
\begin{eqnarray}
  \tilde g  &\to& q \bar q \tilde Z_i\ (i=1,2,3,4),\ q \bar q' \tilde W_j\
                 (j=1,2),\ g \tilde Z_1 \;, \label{heavy a}\\
  \tilde q_L &\to& q \tilde Z_i\ (i=1,2,3,4),\ q' \tilde Wj\ (j=1,2)\;,
\label{heavy b}\\
  \tilde q_R &\to& q \tilde Z_i\ (i=1,2,3,4)\;. \label{heavy c}
\end{eqnarray}
Some decays go via loops (e.g.\ $\tilde g\to g \tilde Z_1)$; we have not
attempted an exhaustive listing here.
Figure~3 shows how gluino-to-heavy-gaugino branching fractions increase
with $m_{\tilde g}$ in a particular example
(with $m_{\tilde g} < m_{\tilde q}$)~\cite{bbkt}.

\medskip
\begin{center}
{\small Fig.~3. Example of gluino decay branchings versus mass~\cite{bbkt}.}
\end{center}

\bigskip

The heavier gauginos then decay too:
\begin{eqnarray}
  \tilde W_j &\to& Z \tilde W_k,\, W \tilde Z_i,\, H_i^0 \tilde W_k,\,
                 H^\pm \tilde Z_i,\,f\tilde f   \;, \\
  \tilde Z_i &\to& Z \tilde Z_k,\, W \tilde W_j,\, H_i^0 \tilde Z_k,\,
                 H^\pm \tilde W_k,\, f\tilde f' \;.   \label{Ztwid decay}
\end{eqnarray}
Here it is understood that final $W$ or $Z$ may be off-shell and
materialize as fermion-antifermion pairs; also $Z$ may be replaced
by $\gamma$.  In practice, chargino decays are usually dominated by
$W$-exchange transitions; neutralino decays are often dominated
by sfermion exchanges  because the $\tilde Z_2 \tilde Z_1 Z$
coupling is small.  To combine the complicated production and cascade
possibilities systematically, all these channels have been incorporated in
the ISAJET~7.0 Monte Carlo package called ISASUSY~\cite{bppt}.

   These multibranch cascade decays lead to higher-multiplicity final
states in which the LSPs $\tilde Z_1$  carry a much smaller share of the
available energy, so $\overlay/p_T$ is smaller and less distinctive,
making detection via $\overlay/p_T$ more difficult.
(Recall that leptonic $W$ or $Z$ decays,
$\tau$ decays, plus semileptonic $b$ and $c$ decays, all give background events
with genuine $\overlay/p_T$; measurement uncertainties also contribute fake
$\overlay/p_T$ backgrounds.)  Experimental bounds therefore become weaker when
we take account of cascade decays.  Figure~4 shows typical CDF 90\% CL
limits in the $(m_{\tilde g}, m_{\tilde q})$ plane; the dashed curves are
limits assuming only direct decays (\ref{light a})--(\ref{light b}), while
solid curves are less restrictive limits including cascade decays~(\ref{heavy
a})--(\ref{Ztwid decay}).

\medskip
\begin{center}
\parbox{5.5in}{\small Fig.~4. 1992 CDF limits in the $(m_{\tilde g}, m_{\tilde
q})$ plane, with or without cascade decays, for a typical choice of
parameters~\cite{btw,uacdf}. }
\end{center}
\medskip

Cascade decays also present new opportunities for SUSY detection.
   Same-sign dileptons (SSD) are a very interesting signal~\cite{bkp}, which
arises naturally from $\tilde g \tilde g$  and  $\tilde g \tilde q$  decays
because of the Majorana character of gluinos, with very
little background. Figure~5 gives an example.
Eqs.~(\ref{heavy a})--(\ref{Ztwid decay})  show how a heavy gluino or squark
can decay to a chargino
$\tilde W_j$ and hence, via a real or virtual $W$, to an isolated charged
lepton.  If a gluino is
present it can decay equally into either sign of chargino and lepton
because it is a Majorana fermion.  Hence  $\tilde g \tilde g$  or
$\tilde g \tilde q$  systems can decay to isolated SSD plus jets plus
$\overlay/p_T$.  Cascade decays of $\tilde q\tilde q$ via the heavier
neutralinos $\tilde Z_i$ offer similar possibilites for SSD, since the $\tilde
Z_i$ are also Majorana fermions. Cross sections for the Tevatron are
illustrated in Fig.~6.

\medskip
\begin{center}
{\small Fig.~5. Example of same-sign dilepton in gluino-pair decay.}

\bigskip

{\small Fig.~6. Same-sign dilepton signals at the Tevatron~\cite{baerev}.}
\end{center}

\bigskip

   Genuinely isolated SSD backgrounds come from the production of $WZ$ or
$Wt\bar t$ or $W^+W^+$ (e.g.\ $uu \to ddW^+W^+$ by gluon exchange), with cross
sections of order  $\alpha_2^2$   or   $\alpha_2\alpha_3^2$    or
$\alpha_2^2\alpha_3^2$  compared to $\alpha_3^2$ for gluino pair production,
so we expect to control them with suitable cuts.  Very large   $b \bar b$
production gives SSD via semileptonic $b$-decays plus $B$-$\bar B$ mixing, and
also via combined $b\to c\to s \,\ell^+ \nu$   and  $\bar b\to \bar c\,\ell^+
\nu$ decays, but both leptons are produced in jets and can be suppressed by
stringent isolation criteria.  Also $t\bar t$ gives SSD via $t\to b\, \ell^+
\nu$ and  $\bar t\to \bar b\to \bar c\, \ell^+ \nu$, but the latter lepton is
non-isolated.
So SSD provide a promising SUSY signature.

The Tevatron can also search for trileptons\cite{tev3,lopez}, arising
for example from
%
 $W^* \to \tilde W \tilde Z_2 \to
               (\ell \nu \tilde Z_1)(\ell \bar \ell \tilde Z_1)$.
%
Figure~7 shows predicted trilepton rates at the Tevatron for both
the minimal SU(5) and no-scale flipped SU(5) supergravity models;\cite{lopez}
the flipped SU(5) model tends to give the bigger
values.

   Gluino production rates at SSC/LHC are much higher than at the Tevatron. At
$\sqrt s=40$~TeV, the cross section is
\begin{equation}
\sigma(\tilde g \tilde g) = 10^4,\,700,\, 6\;\mbox{fb\quad for }
 m_{\tilde g} = 0.3,\,1,\,2\rm\; TeV\;.
\end{equation}
Many different SUSY signals have been evaluated, including
$\overlay/p_T + n\,$jets, $\overlay/p_T +{}$SSD, $\overlay/p_T + n\,$isolated
leptons, $\overlay/p_T + {}$one isolated lepton${}+ Z$, $\overlay/p_T + Z$,
$\overlay/p_T + Z + Z$.
SSC  signals from  $\tilde g \tilde g$
production are shown versus $m_{\tilde g}$ in Fig.~8 for two scenarios,
after various cuts;  the labels 3,4,5 refer to numbers of isolated
leptons~\cite{btw}.

Sufficiently heavy gluinos can also decay copiously to
$t$-quarks~\cite{btw,bps}:
\begin{equation}
\tilde g \to t \bar t\tilde Z_i , t \bar b \tilde W^-, b \bar t
\tilde W^+ \;.
\end{equation}
Then $t\to bW$  decay leads to multiple $W$ production.  For  a
gluino of mass 1.5~TeV,  the  $\tilde g\to W,\, WW,\, WWZ,\, WWWW$ branching
fractions are typically of order 30\%, 30\%,{\parfillskip0pt\par}

\medskip
\begin{center}
\parbox{5.5in}{\small Fig.~7. Predicted Tevatron trilepton events per 100
pb$^{-1}$ for minimal SU(5) and no-scale flipped SU(5) supergravity solutions
(each represented by a point)~\cite{lopez}.}
\end{center}
\medskip

\begin{center}
{\small Fig.~8. SSC cross sections for various SUSY signals, after
cuts~\cite{btw}.}
\end{center}

\noindent
 6\%, 6\%, respectively. Figure~9 illustrates SSC cross sections for multi-$W$
production via gluino pair
decays (assuming $m_{\tilde g} < m_{\tilde q}$).  We see that
for $m_{\tilde g}\sim 1$~TeV the SUSY rate for $4W$ production can greatly
exceed the dominant SM $4t\to 4W$ mode, offering yet another signal for
SUSY~\cite{bps}.

\medskip
\begin{center}
{\small Fig.~9. Typical SSC rates for for $\tilde g \tilde g \to{}$multi-$W$
states~\cite{bps}.}
\end{center}

To summarize this section:
\begin{enumerate}
\addtolength{\itemsep}{-.1in}

\item Experimental SUSY particle searches have hitherto been based
largely on $\overlay/p_T$ signals.  But for $m_{\tilde g},\,m_{\tilde q}
> 50$~GeV cascade decays become important; they weaken the simple
$\overlay/p_T$ signals but provide new signals such as same-sign dileptons and
trileptons, which will be pursued at the Tevatron.

\item For even heavier squarks and gluinos, the cascade decays dominate
completely and provide further exotic (multi-$W,Z$ and multi-lepton)
signatures, which will be pursued at the SSC and LHC.
They would find spectacular events, containing several hard jets from the
primary decay plus many hard isolated leptons ($10^3$--$10^4$
events per year with 3--4 such leptons), sometimes having
multiple $t$ and $b$ hadrons in the chain, with little conventional
background.

\item Gluinos and squarks in the expected mass range
will not escape detection.

\end{enumerate}

\section{SUSY Higgs Phenomenology}

In minimal SUSY, two Higgs doublets $H_1$ and $H_2$ are needed to cancel
anomalies and at the same time give masses to both up- and down-type
quarks\cite{hhg}. Their vevs are $v_1=v\cos\beta$ and $v_2=v\sin\beta$ where
$v = 246$~GeV is the SM vev and $\tan\beta = v_2/v_1$ parameterizes their ratio
($0 \leq \beta \leq \pi/2$). There are therefore 5 physical scalar states: $h$
and $H$ (neutral CP-even with $m_h<m_H$), $A$ (neutral CP-odd) and $H^\pm$. At
tree level the scalar masses and couplings and an $h$-$H$ mixing angle $\alpha$
are all determined by two parameters, conveniently chosen to be $m_A$ and
$\tan\beta$. At tree level the masses obey $m_h\le M_Z,m_A; m_H\ge M_Z,m_A;
m_{H^\pm}\ge M_W,m_A$.

Radiative corrections can be big, however~\cite{rad}. The most important new
parameters entering here are the $t$ and $\tilde t$ masses; we neglect for
simplicity some other parameters related to squark mixing. One-loop corrections
give $h$ and $H$ mass shifts of order $\delta m^2\sim G_F\,m_t^4\ln(m_{\tilde
t}/m_t)$, arising from incomplete cancellation of $t$ and $\tilde t$ loops. The
$h$ and $H$ mass bounds get shifted up and for the typical case $m_t=150$~GeV,
$m_{\tilde t}=1$~TeV (which we usually assume) we get
\begin{equation}
m_h \ltap 115\gev \ltap m_H \,.
\end{equation}
There are also corrections to cubic $hAA,\,HAA,\,Hhh$ couplings, to $h$-$H$
mixing, and smaller corrections to the $H^\pm$ mass. Figure~10 illustrates the
dependence of $m_h$ and $m_H$ on $m_A$ and $\tan\beta$.

\medskip
\begin{center}
\parbox{5.5in}{\small Fig.~10. $h$ and $H$ mass contours in the $(m_A, \tan
\beta)$ plane  for (a)~$m_t=150$~GeV and (b)~$m_t=200$~GeV, with $m_{\tilde t}
= 1$~TeV.}
\end{center}

In the next-to-minimal SUSY model (NMSSM), one Higgs singlet is added\cite{hhg}
giving 7 bosons $h^0,H_1^0,H_2^0,A_1^0,A_2^0,H^{\pm}$, with
6 parameters at tree level (because there are now more terms in the
general scalar potential).  This leads to a richer mass spectrum with
looser limits.  It is instructive to see what happens to the upper
bound on the lightest scalar mass $m_h$, as constraints on the Higgs sector,
the gauge sector and fermion sector are progressively relaxed while
requiring perturbativity up to scale $\Lambda = 10^{16}$~GeV and a SUSY
scale of order $M_{\rm SUSY} \sim 1$~TeV~\cite{relax}:
\[
\vbox{\tabskip1.5em
\halign{#\hfil&$#$\hfil&#\hfil\cr
within MSSM:&  m_h \leq M_Z&       [tree level]\cr
            & m_h \ltap 115\gev&   [rad.corr., $m_t=150$~GeV]\cr
            & m_h \ltap 130\gev&   [rad.corr., any $m_t$]\cr
beyond MSSM: & m_h \ltap 145\gev&  [arbitrary gauge singlets]\cr
             & m_h \ltap 155\gev&  [arbitrary Higgs sector]\cr
             & m_h \ltap 180\gev&  [extra matter multiplets]\cr}}
\]


What are the constraints on the principal parameters $m_A$ and $\tan \beta$?

\begin{enumerate}
\addtolength{\itemsep}{-.12in}

\item
 Perturbativity. This condition is a bit subjective. Requiring
the $tbH^{\pm}$ couplings to be less than $g_3(M_Z) \simeq 1.2$
gives $m_t/(500\gev) \ltap \tan \beta \ltap (500\gev)/m_b$~\cite{bhp}.
Requiring instead (two-loop)/(one-loop)${}< 1/4$ in the renormalization
group equations (RGE) gives $\tan \beta \ltap 65$~\cite{bbo}.

\item
 Proton decay. Some models require $\tan \beta < 3$--15~\cite{lopez2},
but a more conservative limit is $\tan \beta < 85$~\cite{hisano}.

\item
 SUSY-GUT models~\cite{ross,rosrob,bbo,lopez,dhr} find solutions
in the range $0.6 \ltap \tan \beta \ltap 65$ (with the upper bound
from perturbativity).

\item
At LEP\,I, all four experiments~\cite{LEP} have searched for the processes
\begin{equation}
e^+e^- \to Z \to Z^*h,Ah \,,
\end{equation}
with $Z^*\to\ell\ell,\nu\nu,jj$ plus $h,A\to\tau\tau,jj$ and $h\to AA$ decay
modes. The absence of signals excludes regions of the $(m_A,\tan\beta)$ plane
\begin{equation}
m_h,m_A\agt20\mbox{--45~GeV (depending on $\tan\beta$ for $\tan\beta\gtap1$)}
\,.
\end{equation}
Null searches for $e^+e^-\to H^+H^-$ also exclude a region with
$\tan\beta<1$~\cite{diaz}.

\item
 Global electroweak analysis with MSSM\cite{fogli} prefers $m_A <{}$a few
100~GeV, but without much statistical weight.

\item
 To cancel divergences and to remain within the GUT scale,
$m_A \ltap 1$--2~GeV.  Combining (a)--(f), there is a general consensus that
\begin{eqnarray}
      &0.6 \ltap \tan \beta \ltap 65   \,, \label{tanbetarange}\\
%
     &20\gev  \ltap  m_A  \ltap  \mbox{1--2 TeV} \,.  \label{m_A range}
\end{eqnarray}

\end{enumerate}

\noindent
There is also indirect evidence from various quarters:

\begin{enumerate}\setcounter{enumi}{6}
\addtolength{\itemsep}{-.12in}

\item
The measured $B \to \tau \nu X$ branching fraction\cite{btaunu} is
not far from the SM pred-\break
iction and constrains possible virtual $H^{\pm}$
contributions, giving\cite{isidori} $\tan \beta <\break
 0.54 m_{H^\pm}/$(1~GeV); this is weaker than
Eqs.~(\ref{tanbetarange})--(\ref{m_A range}) however.

\item
 The experimental bound\cite{cleo}  $B(b \to s \gamma)< 5.4\times 10^{-4}$
is close to the SM prediction, based on $W+t$ loop calculations, setting
stringent limits on non-SM contributions.  If just $H^{\pm}+t$ loops are
added, an important part of the $(m_A,\tan\beta)$ plane is excluded\cite{bsg}.
However, if spartners are relatively light, additional
$\tilde t + \tilde W_i$ loops may contribute significantly with either
sign\cite{bertolini}. We need more information about the SUSY spectrum
to exploit this bound.

\end{enumerate}

   Future direct searches for MSSM  Higgs bosons rely on future colliders
(apart from marginal improvements at LEP\,I as higher luminosity accumulates).
Several groups\cite{baer,gunion,kunszt,bcps} have studied the prospects
for discovering the different bosons at LEP\,II and SSC/LHC; their results
broadly agree.

LEP\,II will have higher energy $\sqrt s \ltap 200$~GeV, and should be
able to discover $h$ through the range $m_h \ltap 90$~GeV, but will not be
able to cover the full range of possible $m_h$.  $A$ might be light enough
to discover, but $H$ and $H^{\pm}$ are likely to be too heavy. The parameter
sector where both  $m_A \gtap 90$~GeV and  $\tan \beta \gtap 5$ will not be
accessible --- i.e.\ will give no LEP\,II Higgs signals for $m_t\sim150$~GeV.

Searches for neutral scalars at SSC and LHC will primarily be analogous to SM
Higgs searches (see Figs.~11--13):

\begingroup
\renewcommand{\labelenumi}{(\roman{enumi})}
\begin{enumerate}
\addtolength{\itemsep}{-.1in}

\item untagged $\gamma\gamma$ signals from $pp\to(h,H,A)\to\gamma\gamma$ via
top quark loops;

\item lepton-tagged $\gamma\gamma$ signals from $pp\to(h,H,A)\to\gamma\gamma$
plus associated $t\bar t$ or $W$, with leptons from $t\to W\to \ell\nu$ or
$W\to\ell\nu$ decays;

\item ``gold-plated'' four-lepton signals from $pp\to(h,H)\to ZZ$ or
$Z^*Z\to\ell^+\ell^+\ell^-\ell^-$.
\end{enumerate}

\endgroup

\noindent
Though qualitatively similar to SM signals, these will generally be smaller due
to the different couplings that depend on $\beta$ and $\alpha$.

\medskip
\begin{center}
{\small Fig.~11. Typical diagram for untagged Higgs${}\to\gamma\gamma$
signals.}

\medskip
{\small Fig.~12. Typical diagrams for lepton-tagged Higgs${}\to\gamma\gamma$
signals.}

\medskip

{\small Fig.~13. Typical diagrams for ``gold-plated'' four-lepton Higgs
signals.}
\end{center}

\medskip
For charged Higgs scalars, the only copious hadroproduction source is top
production with $t\to bH^+$ decay (that requires $m_{H^\pm}<m_t-m_b$). The
subsequent $H^+\to c\bar s,\nu\bar\tau$ decays are most readily detected in the
$\tau\nu$ channel (favored for $\tan\beta>1$), with $\tau\to\pi\nu$ decay
(Fig.~14).

\medskip
\begin{center}
{\small Fig.~14. Typical diagram for $\tau$ signals from top decay via
charged-Higgs modes.}
\end{center}
\medskip

\noindent
SM $t$-decays give equal probabilities for $e,\nu,\tau$ leptons via $t\to bW\to
b(e,\mu,\tau)\nu$, but the non-standard $t\to bH^+\to b\tau\nu$ leads to
characteristic excess of $\tau$. The strategy is to tag one top quark via
standard $t\to bW\to b\ell\nu$ decay and to study the $\tau/\ell$ ratio in the
associated top quark decay ($\ell=e$ or $\mu$).

New non-SM complications must now be taken into account, however,
especially for the neutral MSSM Higgs boson signals.   New decay channels
open in addition to the main SM modes $h,H \to WW,ZZ$ and  $h,H,A \to t\bar t,
b \bar b, c \bar c, \tau \bar \tau$  (and also $h,H,A \to \gamma \gamma$
via loops).

\begin{enumerate}
\addtolength{\itemsep}{-.1in}

\item  Firstly there are decays to other Higgs scalars,  $h \to AA,\
A \to Zh,\ H \to hh$ or $AA$ or $AZ$;  the regions where these channels are
open are shown in Fig.~15, for $m_t = 150$~GeV. In particular $H \to hh$ is
allowed and generally dominates (suppressing all the usual SM signatures)
everywhere except in the shaded region or near the line of coupling zeros
labelled $f_h = 0$.  We already mentioned $h\to AA $ in passing; it changes the
$h$ signals in a small region of the LEP\,I searches.

\medskip
\begin{center}
{\small Fig.~15. Allowed regions for decays to other Higgses\cite{bcps}.}
\end{center}
\medskip

\item Secondly, there are decays to SUSY particles, especially the invisible
mode $h \to\tilde Z_1 \tilde Z_1$;  these have usually been ignored but
could be significant\cite{baer2}. They could give new signals
and they could suppress old signals.  Figure~16 shows contours of $B(h \to
\mbox{SUSY, mostly }\tilde Z_1 \tilde Z_1)$ versus $\mu$ and $\tan \beta$,
the region excluded by LEP\,I SUSY searches being shaded (see Section 2);
there are small regions where SUSY modes could rise to 70\%--90\%, severely
suppressing standard $h$ signals, but they are within the reach of LEP\,II SUSY
searches (dashed lines) so we shall know.  Here we shall mostly ignore SUSY
modes, arguing that they may dilute other Higgs signals but will probably not
efface them.

\medskip
\begin{center}
{\small Fig.~16. Contours of $B(H\to{}$SUSY modes) versus $\mu$ and
$\tan\beta$\cite{baer2}.}
\end{center}

\end{enumerate}

\bigskip

Returning to MSSM Higgs searches at SSC/LHC\cite{baer,gunion,kunszt,bcps},
Figs.~17(a) and 17(b) show typical limits of detectability for untagged and
lepton-tagged $\gamma\gamma$ signals at SSC, assuming luminosities ${\cal
L}=20\,\rm fb^{-1}$ (two years of running) and $m_t=150$~GeV. Figure~17(c)
shows a similar limit for the $H\to4\ell$ search (no $h\to4\ell$ signal is
detectable).  Figure~17(d) shows typical limits for detecting the $t\to
H^+\to{}$excess $\tau$ signal; here the value of $m_t$ is critical, since only
the range $m_{H^+}<m_t-m_b$ can contribute at all. Putting all these discovery
regions together with the LEP\,I and LEP\,II regions, we see that very
considerable coverage of the $(m_A,\tan\beta)$ plane can be expected --- but
there still remains a small inaccessible region; see Fig.~18.
For $m_t=120$~GeV the inaccessible region is larger, for $m_t=200$~GeV it is
smaller.

\medskip
\begin{center}
\parbox{5.5in}{\small Fig.~17. Typical detectability limits for (a) untagged
$h,H,A\to\gamma\gamma$, (b)~lepton-tagged $h,H\to\gamma\gamma$,
(c)~$H\to4\ell$, (d)~$H^\pm\to\tau\nu$ signals at the SSC\cite{bcps}.}

\bigskip

\parbox{5.5in}{\small Fig.~18. Combined LEP and SSC discovery regions for
$m_t=150$~GeV from Ref.~33; 
similar results are obtained by other groups~\cite{baer,gunion,kunszt}.}

\end{center}
\medskip

Emphasizing first the positive side, Fig.~19 shows how many of the MSSM scalars
$h,H,A,H^\pm$ would be detectable, in various regions of the $(m_A,\tan\beta)$
plane. In many regions two or more different scalars could be discovered, but
for large $m_A$ only $h$ would be discoverable; in the latter region, the $h$
couplings all reduce to SM couplings, the other scalars become very heavy and
approximately degenerate, and the MSSM essentially behaves like the SM.

\medskip
\begin{center}
{\small Fig.~19.  How many MSSM Higgs bosons may be discovered (from Ref.~33).}
\end{center}
\medskip

Turning now to the negative side of Fig.~18, the inaccessible region,
Fig.~20 shows that different groups broadly agree on the boundary.
This region corresponds to intermediate masses for all Higgses:
$m_h \sim 80$--115~GeV, $m_A \sim 100$--160 GeV, $m_H \sim 120$--160~GeV,
$m_{H\pm} \sim 120$--160 GeV.  Why are none detectable?  What went wrong?
It seems we are just unlucky here: the $htt$ and $hWW$
couplings that control $h \to \gamma \gamma$ become weaker; the $Att$
coupling that controls both $A$ production and two-photon decay gets
weak; $H \to hh$ competition suppresses the $H$ signals; $H^{\pm}$ is too
heavy and $t \to b H$ is suppressed.  Next, is there any indirect way to
exclude this inaccessible region?  One possibility is to understand
and exploit the $B(b \to s \gamma)$ bound better: see above.  We note also
that if $m_t =200$~GeV (instead of 150~GeV as in Fig.~18), the region is
much smaller.  Another possibility is to derive stronger parameter
constraints from SUSY-GUT models, as we now discuss.

\medskip
\begin{center}

{\small Fig.~20. Different groups agree about inaccessible region.}
\end{center}
\medskip

  In minimal SUSY-GUT models with $M_{\rm SUSY} \ltap 1$~TeV, the RGE
have a solution dominated by the infrared fixed point of the top
Yukawa coupling $\lambda_t$\cite{pendleton}.
For large $m_t$,
$\lambda_t$ is plausibly large at the GUT scale $M_G$, in which case
it evolves rapidly toward an infrared fixed point at low mass
scales\cite{bbo,dhr,pendleton,ramond,knowles,pokorski}, according to
the one-loop renormalization group equation
\begin{equation}
{{d\lambda _t}\over {dt}}={{\lambda _t}\over {16\pi ^2}}\left [
-\sum c_ig_i^2+6\lambda _t^2+\lambda _b^2\right ]\;,\label{lambdat}
\end{equation}
with $c_1=13/15$, $c_2=3$, $c_3=16/3$. Thus $\lambda_t$ evolves toward a fixed
point close to
where the quantity in square brackets in Eq.~(\ref{lambdat}) vanishes.
Then the known gauge couplings determine the running
mass $m_t(m_t)=\lambda_t(m_t)v\sin\beta/\sqrt2$ and hence the pole mass
$m_t({\rm pole})=m_t(m_t)\left[1+{4\over3\pi}\alpha_s(m_t)\right]$; two-loop
evaluations\cite{bbo} give
\begin{equation}
m_t({\rm pole})\simeq(200\gev)\sin\beta \,,  \label{mtpole}
\end{equation}
If $\lambda_t(M_G)$ is below the fixed point, its convergence to the fixed
point is more gradual and Eq.~(\ref{mtpole}) does not necessarily apply.
But in practice large $\lambda_t(M_G)$ is favored in many SUSY-GUT solutions;
large $\lambda _t(M_G)$ facilitates $\lambda_b(M_G)=\lambda_\tau(M_G)$
Yukawa unification\cite{CEG} and allows intricate relationships between fermion
masses and mixings. It is therefore interesting to pursue the phenomenological
implication of Eq.~(\ref{mtpole})~\cite{bbop}.

Figure 21 shows how $\lambda_t(M_G)$ and $\lambda_b(M_G)$ values
relate to $m_t(\rm pole)$ and
$\tan\beta$ in SUSY-GUT solutions with Yukawa unification; the lower (upper)
shaded
branches contains the $m_t\ (m_b)$ fixed-point solutions. There is a small
region at the upper right where both fixed point solutions are simultaneously
satisfied. Figure~22 shows that
the $m_t$ fixed-point behavior is insensitive to GUT threshold corrections
$\alt 10\%$ in the $\lambda_b/\lambda_\tau$ ratio. The
sensitivity of the fixed point to threshold corrections is decreased
for larger values of $\alpha _s(M_Z)$ where the solutions tend to have a
stronger fixed point character, as indicated by Eq.~(\ref{lambdat}). The
perturbative limits of the Yukawa couplings near their Landau poles are
shown in Fig.~22(a) as the dashed lines $\lambda _t^G = 3.3$ and
$\lambda _b^G = 3.1$.

\medskip
\begin{center}
\parbox[c]{5.5in}{\small Fig.~21: Contours of constant Yukawa couplings
$\lambda _i^G=\lambda _i(M_G^{})$
at the GUT scale in the ($m_t^{\rm pole}, \tan \beta$) plane,
obtained from solutions to the RGE with
$\lambda _{\tau}^G=\lambda _b^G$ unification imposed.
The GUT scale Yukawa coupling contours are close together for
large $\lambda ^G$. The
fixed points describe the values of the Yukawa couplings at
the electroweak scale for $\lambda _t^G\agt 1$ and
$\lambda _b^G\agt 1$.}

\end{center}

\medskip

\begin{center}
\parbox{5.75in}{\small Fig.~22: RGE results for
$\alpha _s(M_Z^{})=0.118$ with the boundary condition $m_b(m_b)=4.25$
GeV.
(a) GUT threshold corrections to Yukawa coupling unification.
The solutions strongly exhibit
a fixed point nature, for threshold
corrections $\alt 10\%$.
Taking a larger supersymmetric threshold $M_{SUSY}^{}$
or increasing $\alpha _s(M_Z)$
moves the curves to the right, so that the fixed point condition becomes
stronger.
(b)~Evolution of the top quark Yukawa coupling for
$\tan \beta =1$. The dashed line indicates
${{d\lambda _t}\over {dt}}=0$ which gives an
approximation to the electroweak scale value of $m_t$ with accuracy of
order~10\%.}

\end{center}
\medskip

There are immediate implications for phenomenology\cite{bbop}.  If
$m_t\alt160$~GeV, Eq.~(\ref{mtpole}) constrains
$\tan\beta$ to values near 1, where $h$ is relatively light (recall the
tree-level relation $m_h<M_Z|\cos 2\beta |$) and the couplings of $h$ are close
to those of a Standard Model Higgs boson.
LEP Higgs searches\cite{LEP,lepsm} exclude a region of the
$(m_A,\tan\beta)$ plane shown in Fig.~23(a): this translates to forbidden
regions in $(m_h,\tan\beta)$ in Fig.~23(b). We see that the fixed-point
condition
predicts $m_t\agt 130$~GeV, $m_h\agt60$~GeV, $m_A\agt70$~GeV; correspondingly
$m_{H^\pm}\agt105$~GeV,  $m_H\agt140$~GeV. If in fact $m_t\alt160$~GeV, then
$m_h\alt85$~GeV as shown in Fig.~23, and $h$ will be discoverable at LEP\,II
(but none of the
other Higgs bosons will). The discovery limits at SSC/LHC (taken here from
Ref.~33) 
are shown in Fig.~24; we see that each of the five Higgs bosons might be
discoverable there, but not all at once, and possibly none of them at all.

\medskip
\begin{center}
\parbox{5.5in}{\small Fig.~23: $m_t$ fixed-point solution regions allowed by
the LEP\,I data: (a)~in the $(m_A, \tan \beta )$ plane, (b)~in the $(m_h, \tan
\beta )$ plane. The top quark masses are $m_t({\rm pole})$, correlated to
   $\tan \beta $ by Eq.~(\ref{mtpole}).}

\bigskip

\parbox[c]{5.5in}{\small Fig.~24: SSC/LHC signal detectability regions,
compared with the LEP\,I allowed region of $m_t$ fixed-point solutions from
Fig.~3(a) and the
probable reach of LEP\,II. The top quark masses are $m_t({\rm pole})$.}

\end{center}

\medskip

Finally we may ask what a future $e^+e^-$ collider could do. We have seen that
part of the MSSM parameter space is inaccessible to $e^+e^-$ collisions at
$\sqrt s=200$~GeV, $\L=500\pb^{-1}$, for $m_t=150$~GeV and $m_{\tilde
t}=1$~TeV. But a possible future linear collider with higher energy and
luminosity could in principle cover the full parameter space. In is interesting
to know what are the minimum $s$ and $\L$ requirements for complete coverage,
for given $m_t$. This question was answered in Ref.~42, 
based on the conservative assumption that only the channels
$e^+e^-\to(Zh,Ah,ZH,AH)\to\tau\tau jj$ would be searched, with no special
tagging. The results are shown in Fig.~25. We have estimated that including all
$Z\to\ell\ell,\nu\nu,jj$ and $h,H,A\to bb,\tau\tau$ decay channels plus
efficient $b$-tagging could increase the net signal $S$ by a factor~6 and the
net background $B$ by a factor~4, approximately; this would increase the
statistical significance $S/\sqrt B$ by a factor~3 and hence reduce the
luminosity requirement by a factor~9 or so. In this optimistic scenario, the
luminosity scale in Fig.~25 would be reduced by an order of magnitude.

\medskip
\begin{center}
\parbox[c]{5.5in}{\small Fig.~25. Minimal requirements for a ``no-lose'' MSSM
Higgs search at a future $e^+e^-$ collider based on $\tau\tau jj$ signals
alone. Curves of minimal $(\sqrt s,\L)$ pairings are shown for $m_t=120$, 150,
200~GeV; the no-lose region for $m_t=150$~GeV is unshaded~\cite{nolose}.}
\end{center}
\medskip

To summarize this Section:

\begin{enumerate}
\addtolength{\itemsep}{-.1in}

\item The MSSM Higgs spectrum is richer but in some ways more elusive than the
SM case.

\item At least one light scalar is expected.

\item As $m_A\to\infty$ this light scalar behaves like the SM scalar while the
others become heavy and degenerate.

\item LEP\,I, LEP\,II and SSC/LHC will give extensive but not quite complete
coverage of the MSSM parameter space.

\item For some parameter regions, several different scalars are detectable, but
usually one or more remain undetectable.

\item The $b\to s\gamma$ bound has the potential to exclude large areas of
parameter space (possibly including the inaccessible region) but is presently
subject to some uncertainty.

\item A higher-energy $e^+e^-$ collider could cover the whole MSSM parameter
space, discovering at least the lightest scalar $h$.

\item $m_t$ fixed-point solutions in SUSY-GUT
                         models are theoretically attractive and also
                         strongly constrain the phenomenology; they
                         predict $m_t \gtap 130$~GeV;  if $m_t \ltap 160$
                         GeV, they exclude the inaccessible region and
                         predict that $h$ will be discovered at LEP.

\end{enumerate}


\noindent{\bf Acknowledgments:} {\small
This work was supported in part by the University of Wisconsin Research
Committee with funds granted by
the Wisconsin Alumni Research Foundation, in part by the U.S.~Department of
Energy under contract no.~DE-AC02-76ER00881, and in part by the Texas National
Laboratory Research Commission under grant no.~RGFY93-221.}

\end{document}